\documentclass[twocolappendix,11pt,a4paper]{emulateapj}

\usepackage{psfrag}
\usepackage{psfig}
\usepackage{pspicture}
\usepackage{graphicx}
\usepackage[export]{adjustbox}
\usepackage{amssymb,amsmath}
\usepackage{natbib}

\begin{document}

\title {Dynamics of Star-forming Galaxies in a Massive Structure at \lowercase{$z\sim$} 2.2: Evidence for Galaxy Harassment in high-$z$ Environments}

\author{
Behnam Darvish\altaffilmark{1,2},
Nima Chartab\altaffilmark{3},
Zahra Sattari\altaffilmark{3},
Sina Taamoli\altaffilmark{1},
Irene Shivaei\altaffilmark{4},
Nick Scoville\altaffilmark{5},
Shoubaneh Hemmati\altaffilmark{3},
David Sanders\altaffilmark{6}
}

\setcounter{footnote}{0}

\altaffiltext{1}{Department of Physics and Astronomy, University of California, Riverside, 900 University Ave, Riverside, CA 92521, USA; email: bdarv001@ucr.edu}
\altaffiltext{2}{Eureka Scientific, 2452 Delmer St, Oakland, CA 94602}
\altaffiltext{3}{IPAC, Mail Code 314-6, California Institute of Technology, 1200 East California Boulevard, Pasadena, CA 91125, USA}
\altaffiltext{4}{Steward Observatory, University of Arizona, Tucson, AZ 85721, USA}
\altaffiltext{5}{Cahill Center for Astrophysics, California Institute of Technology, 1216 East California Boulevard, Pasadena, CA 91125,
USA}
\altaffiltext{6}{Institute for Astronomy, 2680 Woodlawn Dr., Honolulu, HI
96822, USA}

\begin{abstract}

We spectroscopically confirm a new protocluster in the COSMOS field at $z$=2.24430 with Keck/MOSFIRE, dubbed CC2.2B, which is in the immediate vicinity of CC2.2A protocluster, originally presented in \cite{Darvish20}. CC2.2B and CC2.2A centroids are separated by $\sim$5.5 Mpc(angular) and $\sim$16 comoving Mpc(radial). CC2.2B and CC2.2A have similar properties, with CC2.2B having a line-of-sight velocity dispersion and estimated total mass of $\sigma_{los}$=693$\pm$65 km s$^{-1}$ and $M_{total}$=($\sim$2-3)$\times$10$^{14}$ $M_{\odot}$, respectively. These two similar overdensities are likely still in the merging process and will likely collapse into a more massive structure at lower redshifts. We combine CC2.2A and CC2.2B data to investigate the role of high-$z$ protocluster environments on the dynamics of star-forming (SF) galaxies compared to a similarly selected field sample. We find that on average, protocluster SF galaxies have $\sim$0.1 dex (at $\sim$1.8$\sigma$ significance) lower gas velocity dispersions, $\sim$0.2 dex (at $\sim$2.2$\sigma$ significance) lower dynamical masses, and $\sim$0.2 dex lower dynamical-to-stellar mass ratio than the field SF galaxies. We argue that galaxy harassment and galaxy-galaxy interactions can potentially explain these differences. We also find a factor of $\sim$2-3 lower scatter around the mean $\sigma$-$M_{*}$, $M_{dyn}$-$M_{*}$, and $M_{dyn}$/$M_{*}$ vs. $M_{*}$ relations for protocluster SF galaxies than the field. This could be due to a more uniform formation for protocluster galaxies than their field counterparts. Our results have potential implications for the physics of preprocessing in early environments.    
    
\end{abstract}


\section{Introduction} \label{int}

At low redshifts, it is well known that many properties of galaxies, such as their morphology, stellar mass, star formation rate, gas content, color, and metallicity strongly depend on their host environment (e.g., \citealp{Dressler80,Peng10,Darvish16}). However, at high redshifts ($z\gtrsim$2), there is not much consensus on the relation between the environment and galaxy properties. For example, conflicting results have been found for the relation between dense environment and galaxies SFR (e.g., \citealp{Tran15,Darvish16,Chartab20}), gas-phase metallicity (e.g., \citealp{Kacprzak15,Shimakawa15,Sattari21}), gas content (e.g., \citealp{Lee17,Darvish18b,Hayashi18}) and so on. A likely reason for these seemingly inconsistent results is the small sample size, along with other potential factors (see e.g., \citealp{Darvish20}).
  
Simulations have shown that some physical mechanisms with thermal, gravitational, or hydrodynamical origins can effectively act in dense environments and shape the properties of galaxies. These include ram-pressure stripping \citep{Gunn72}, thermal evaporation \citep{Cowie77}, galaxy-galaxy interactions \citep{Merritt83}, galaxy mergers \citep{Lin10}, galaxy harassment \citep{Moore98}, viscous stripping \citep{Nulsen82}, strangulation \citep{Larson80}, galaxy-cluster tidal interactions \citep{Merritt84}, among others. These can potentially result in galaxies ``losing'' their mass, gas, gas reservoir or ``speeding up'' their gas consumption, transforming them from active star-forming galaxies to passive red systems. These physical mechanisms operate at different effective timescales and their potential effects on galaxies may evolve with cosmic time (e.g., \citealp{Wetzel13,Darvish16}). Therefore, observational studies of properties of galaxies in different environments and redshifts provide a crucial benchmark to pinpoint what physical mechanisms are at play over the evolutionary history of galaxies.      

Dynamics of galaxies is one of their fundamental and observationally derivable parameters. Kinematical measures for galaxies are incorporated in some of the fundamental scaling relations in observational astronomy. These include the rotational velocity of spiral galaxies represented in the Tully-Fisher relation \citep{Tully77} and the stellar velocity dispersion of elliptical systems depicted in the Fundamental Plane \citep{Dressler87} and its projection as the Faber-Jackson relation \citep{Faber76}. These empirically-derived relations can be used to obtain and calibrate distances and more importantly, they provide crucial observational insights for the theoretical models of galaxy formation and evolution.  

While most kinematical studies to $z\sim$1, such as those of the Tully–Fisher, Faber-Jackson or Fundamental Plane relations have found no or at best very weak environmental dependence (e.g., \citealp{Nakamura06,Jaffe11,Mocz12,Rawle13,Darvish15b}), there are still some studies that have found evidence for the dependence of the dynamical measures of galaxies on their host environment. For example, There is evidence for the environmental dependence of the slope \citep{Park07} or intrinsic scatter \citep{Focardi12} of the Faber-Jackson relation in the local universe.
 
The picture is even less coherent at higher redshifts and there are limited studies on the potential role of galaxy environment on its kinematics at $z\gtrsim$2. This is partially attributed to a small sample of spectroscopically-confirmed structures at high-$z$ that would have large samples of high quality data to allow such analysis. This sets the needs for conducting such studies at high redshifts. As a first attempt, \cite{Alcorn16} used the integrated line-of-sight velocity dispersion of H$\alpha$ emission line as a proxy for galaxy kinematics using a sample of cluster and field star-forming galaxies at $z\sim$ 2 and found no significant environmental effects.

\cite{Darvish20} discovered the presence of a potentially multi-component large-scale structure (LSS) in the COSMOS field at $z\sim$2.2. They spectroscopically confirmed one of its overdensities as a $\sim$10$^{14} M_{\odot}$ protocluster at $z\sim$2.2, dubbed CC2.2, that likely developed into a Coma-cluster like structure at present time. Followup studies using the CC2.2 data and a control sample of field galaxies at similar redshifts showed that the dust properties of galaxies (through the IRX-$\beta$ relation) are independent of their host environment \citep{Shivaei20}. The data were also used to find tentative evidence ($\sim$2.5$\sigma$significance) for a gas-phase metallicity deficiency ($\sim$0.1 dex) in massive ($\sim$10$^{10}$-10$^{11}$ $M_{\odot}$) protocluster galaxies compared to the field \citep{Sattari21}. In this paper, we first provide new spectroscopic confirmation for another overdensity in the $z\sim$2.2 LSS in the proximity of CC2.2. We then combine CC2.2 and this work's protocluster data to investigate the potential role of high-$z$ protocluster environments on the kinematics of galaxies using the integrated velocity dispersion of H$\alpha$ or [O{\sc iii}]$\lambda$5007 nebular lines.    

In this paper, we assume a flat $\Lambda$CDM cosmology with $H_{0}$=70 km s$^{-1}$ Mpc$^{-1}$, $\Omega_{m}$=0.3, and $\Omega_{\Lambda}$=0.7. Unless otherwise stated, the transverse cosmological distances are presented as physical distances. The ``physical'' scale at $z$ $\sim$ 2.2 is $\sim$ 0.5 Mpc per arcmin.
         
\section{Super Protocluster Selection} \label{motivation}

\begin{figure*}
\centering
\includegraphics[width=7.0in]{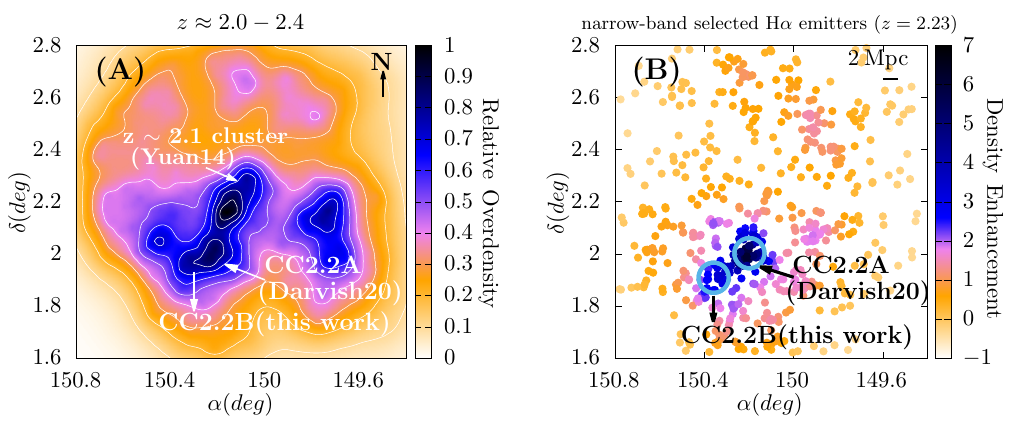}
\caption{
Left panel --- (A) Constructed normalized overdensity map in the COSMOS field at $z$=2.23 within a redshift width of $\Delta z \sim \pm$ 0.2(also see \citealp{Darvish20} for details) using weighted adaptive kernel method of \cite{Darvish15a}. A potentially massive multi-component structure is seen. As labeled, two of these components were already confirmed spectroscopically as a Virgo-like progenitor at $z\sim$2.1 \citep{Yuan14} and a Coma-like progenitor at $z\sim2.2$ \citep{Darvish20}. Right panel --- (B) Distribution of candidate narrow-band selected H$\alpha$ emitters from the HiZELS survey \citep{Sobral13} in the COSMOS field at $z$ $\sim$ 2.23, color-coded by their density enhancement (see \citealp{Darvish20} for details). Both of these maps show the presence of another potential overdensity, labeled as CC2.2B, next to CC2.2A. Followup spectroscopy presented in this paper (Section \ref{observation}) confirm CC2.2B as a new protocluster (Section \ref{property}).
}
\label{fig:map}
\end{figure*}

\cite{Darvish20} introduced the presence of a potentially multi-component massive several Mpc-scale structure seen in photometrically constructed overdensity maps in the COSMOS field at $z$ $\sim$ 2.2. The northern section of this structure had already been confirmed spectroscopically as a protocluster at $z$ $\sim$ 2.1 by \cite{Yuan14}. \cite{Darvish20} spectroscopically confirmed another overdensity in this LSS as a protocluster at $z_{mean}$=2.23224 $\pm$ 0.00101 dubbed CC2.2. CC2.2 is also identifiable as an overdensity of narrow-band selected H$\alpha$ emitting candidates at $z$ $\sim$ 2.2. However, both the photometrically constructed overdensity map as well as the distribution of narrow-band selected H$\alpha$ emitters at $z$ $\sim$ 2.2 show the presence of another potential overdensity adjacent to CC2.2 (see Fig. \ref{fig:map}). From now on, we call CC2.2 as CC2.2A and its potential new protocluster neighbor as CC2.2B. In this paper, we perform follow-up spectroscopic observations with Keck/MOSFIRE targeting CC2.2B and confirm it as a new protocluster (Sections \ref{observation} and \ref{property}). We further combine the data from CC2.2A and CC2.2B and investigate the potential role of high-$z$ protocluster environments on the dynamics of SF galaxies relative to a control sample of similarly selected field galaxies (Sections \ref{vd} and \ref{dm}). 

\section{New Spectroscopic Observations} \label{observation}

We follow the same methodology presented in \cite{Darvish20} to perform spectroscopic sample selection, observations, data reduction and redshift estimation for the potentially new protocluster CC2.2B. We briefly explain the approach here and refer the reader to \cite{Darvish20} for details.

As the primary targets, we use the narrow-band selected H${\alpha}$ emitting candidates from the HiZELS survey \citep{Sobral13} at $z$ $\sim$ 2.23 in the vicinity of CC2.2B. The primary targets are complete down to a stellar mass limit of $\gtrsim$ 10$^{9.7}$ $M_{\odot}$ (see \citealp{Sobral13,Sobral14}). In designing masks for multi-object spectroscopy, we also added fillers to our parent sample. The fillers are selected to be in the proximity of CC2.2B, classified as $NUV-r$ vs. $r-J$ color selected star-forming galaxies, with photometric redshifts in the range 1.7 $<$ $z_{phot}$ $<$ 2.8. These criteria may result in some fillers belonging to the CC2.2B structure as well.

The new observations were conducted on March 10, 2022 using Keck/MOSFIRE near-IR multi-object spectrograph under clear conditions and an average $\sim$ 0.7$^{\prime \prime}$ seeing. We designed two masks and conducted observations in $K$ band, targeting the H$\alpha$ line and other weaker nebular lines. In total, we placed 38 primary targets and fillers on these masks. We used ABBA dithering pattern and observed each mask for a total on-target integration time of 156 and 72 minutes, respectively.  

To reduce the data, we used the MOSFIRE data reduction pipeline \citep{Konidaris19}. The results are cosmic-ray removed, flat-field corrected, sky subtracted, and vacuum wavelength calibrated 2D spectra and their uncertainties per slit. The 1D spectra and the associated uncertainties were later extracted using the optimal extraction algorithm of \cite{Horne86}. 

With the extracted 1D spectra, we obtain redshifts for sources that have at least two $\geqslant$ 3$\sigma$ emission lines by taking the average of the peak of the lines. All of our 20 primary targets yield secure redshifts and 16 emerged as CC2.2B members. Also one filler yields spectroscopic redshift at the expected redshift of CC2.2B. With these data, we confirm CC2.2B as another protocluster in the vicinity of CC2.2A (Section \ref{property}).

\section{Data} \label{data}

\subsection{MOSFIRE NIR Spectroscopic Data} \label{spec}

For the protocluster sample, we combine the data obtained for CC2.2A and CC2.2B to increase the sample size. They all have Keck/MOSFIRE $K$ band observations, with a small fraction of them having $H$ band data as well. The extra $H$ band observations can be used to investigate systematics in measuring the dynamical measures. The data in both bands are located in the COSMOS field with a total on-target integration time of typically $\sim$1-2 hours (see \cite{Darvish20} and Section \ref{observation} here for details).

For the field sample, we compile the data taken mostly in Keck/MOSFIRE $K$ band (with a small fraction in $H$ band) over several years in the COSMOS and UDS fields (PI Nick Scoville). The field sample galaxies have similar selection functions as the protocluster data except that they were put on masks placed randomly in either the COSMOS or UDS fields. Similar to the protocluster data, they are primarily narrow-band selected H$\alpha$ emitting candidates from the HiZELS survey in the COSMOS or UDS fields with fillers selected similarly as those in the protocluster sample (\citealp{Darvish20} and Section \ref{observation}). We also include in the field sample the galaxies originally placed on protocluster masks but turned out to be field SF galaxies. The field sample also has a total on-target exposure time of $\sim$ 1-2 hours.  

We remove sources that are in the final stages of merger which have blended spectra by visually inspecting their spectra and/or near-IR images. We also remove X-ray AGN (using Xray catalogs in the COSMOS AND UDS fields), IR AGN (using \cite{Donley12} criteria), optical AGN (using [NII]/H$\alpha$ $>$ 0.5 if the lines are available) and broad-line AGN (by visually inspecting the spectra) from our sample. 

The final sample consists of 35(0) protocluster star-forming galaxies and 112(43) field SF galaxies in the COSMOS(UDS) field. 162 galaxies have data in $K$ band only, 4 in $H$ band only, and 24 sources have data in both $H$ and $K$ bands. The protocluster sample has a median redsift of $z_{med-clstr}$=2.237 and a median absolute deviation (MAD) of $\Delta z_{MAD-clstr}$=0.003. The median redshift and redshift MAD for the field sample is $z_{med-field}$=2.223 and $\Delta z_{MAD-field}$=0.067, respectively. These values are $z_{med-field-COSMOS}$=2.197 and $\Delta z_{MAD-field-COSMOS}$=0.080 for field galaxies in the COSMOS area only.

We use the spectroscopic data to measure the integrated gas velocity dispersion of the emission lines as a means of their kinematics. This is done by fitting a Gaussian function to the brightest and highest S/N ratio emission line in either $K$- (H$\alpha$) or $H$-band ([O{\sc iii}]$\lambda$5007) after masking other nearby emission lines, such as [N{\sc ii}] lines. The standard deviation of the fitted Gaussian function is used as a proxy for the velocity dispersion (see Section \ref{vd}).    

\subsection{Stellar Masses} \label{m*}

We obtain the stellar masses($M_{*}$) using the publicly available catalogs of COSMOS2015 \citep{Laigle16} and SPLASH-SXDF \citep{Mehta18} for the COSMOS and UDS fields, respectively. In both cases, the stellar masses were extracted using SED fitting to the available photometry in the fields. Both catalogs assumed a Chabrier initial mass function, two attenuation curves, a range of stellar metallicities, and an exponentially declining star formation history (COSMOS2015 used a delayed SFH as well). Details are provided in these papers.

\subsection{HST Imaging Data} \label{image}

To estimate the dynamical mass of galaxies, we need reliable morphological measurements. Hence, we use the morphological measures from the catalog of \cite{Leauthaud07} produced by running SExctractor on HST/ACS I-band F814W photometry in the COSMOS field. As a measure of the galaxy size, we use the circularized effective radius $R_{cir}$ defined as $R_{cir}$=$R_{e}\sqrt{q}$ where $R_{e}$ is the half-light effective radius and $q$ is the axis ratio.   

We note that due to the morphological k-correction, it is more appropriate to use near-IR imaging data for size measurement. However, there is only shallow HST/WFC3 H-band F160W photometric data from the COSMOS-DASH survey that would cover the whole COSMOS field. We matched the H-band morphological catalog of \cite{Cutler22} with our protocluster and field sample and found only 23 common sources. Nonetheless, we compare the $R_{cir}$ sizes measured from the HST I-band and H-band images for this limited number of sources and find a very good agreement. This implies that the I-band size measurement can be utilized reliably in our analysis.

We also note that large-scale HST imaging covering the whole UDS field is not available. Therefore, in our dynamical mass analysis, we only rely on data in the COSMOS field (35 protocluster and 112 field galaxies).   

\section{Results}

\subsection{CC2.2B Properties} \label{property}

We estimate the physical properties of CC2.2B following the methodology of \cite{Darvish20}. CC2.2B members are first determined iteratively using a 3$\sigma$ clipping method until a final mean and standard deviation redshift for the protocluster is obtained. 17 galaxies pass the selection criteria as the CC2.2B members. 

With these member galaxies, we obtain a mean redshift and line-of-sight velocity dispersion (defined as $\sigma_{los}$=c$\sigma_{z}$/(1+$z$) where c is the speed of light) of $z_{mean}$=2.24430$\pm$0.00183 and $\sigma_{los}$=693$\pm$65 km s$^{-1}$, respectively. Following \cite{Darvish20}, to explore the role of small sample size, we randomly select 10 galaxies from our members and recalculate the bootstrapped velocity dispersion of $\sigma_{los}(bootstrap)$=652$\pm$91 km s$^{-1}$. This is in agreement with what we estimate using the full sample.

We define the centeroid of the protocluster as the arithmetic mean of the Cartesian unit vectors representing its members. We use it as the center of CC2.2B at RA=150.358176 (deg) and Dec=+1.908819 (deg). We estimate the radius of the protocluster core ($R_{proj}$) as the distance from the center that encompasses 40\% of the members (corresponding to the weight of a 2D Gaussian distribution within one standard deviation). This gives us $R_{proj}$=0.94$\pm$0.20 Mpc.  

Assuming virialization and spherical symmetry for CC2.2B, we estimate its virial mass from its velocity dispersion and projected core radius as $M_{vir}$=(3$R_{proj}\sigma_{los}^{2}$/G)=(3.2$\pm$0.9)$\times$10$^{14}$ $M_{\odot}$, where G is the gravitational constant. Alternatively, we can estimate the halo mass of CC2.2B ($M_{200}$) assuming virialization, spherical symmetry, and that the halo is a spherical region within which the average density is 200$\rho_{c}(z)$, where $\rho_{c}(z)$ is the critical density of the universe at redshift of $z$. With this approach, we acquire $r_{200}$=$\sqrt{3}\sigma_{los}$/(10$H(z)$) and $M_{200}$=$(\sqrt{3}\sigma_{los})^{3}$/(10G$H(z)$) \citep{Carlberg97}, where $H(z)$ is the Hubble parameter at redshift $z$. For CC2.2B, we obtain $r_{200}$=0.52$\pm$0.05 Mpc and $M_{200}$=(1.7$\pm$0.5)$\times$10$^{14}$ $M_{\odot}$. Finally, using the scaling relation for the simulated clusters presented in \cite{Munari13}, we obtain a total mass of $M_{200}$(scaling)=(0.7$\pm$0.2)$\times$10$^{14}$ $M_{\odot}$. These mass estimates are consistent with one another.  

In table \ref{table1}, we summarize the properties of CC2.2B. For comparison, we also tabulate the same properties for CC2.2A from \cite{Darvish20}. We find that CC2.2B has similar characteristics to CC2.2A. The centers of these two overdensities are only separated by $\sim$5.5 Mpc and $\sim$16 Mpc(comoving) on the plane of the sky and line-of-sight direction, respectively. These two similar overdensities are likely still in the merging process and will likely collapse into a more massive structure at lower redshifts. We note that given the proximity of these two protoclusters, the virialization assumption may not be entirely correct and the dynamical estimates, such as the dynamical masses only provide upper limits and order-of-magnitude estimates. 

\begin{table}
\begin{center}
\caption{CC2.2A and CC2.2B Properties} 
\begin{scriptsize}
\centering
\begin{tabular}{lccc}
\hline
\hline
\noalign{\smallskip}
quantity & CC2.2A & CC2.2B\\
& & \\
\hline
\\
RA(deg)& 150.197509 & 150.358176\\
Dec(deg) & +2.003213 & +1.908819\\
$z_{mean}$ & 2.23224$\pm$0.00101 & 2.24430$\pm$0.00183\\
$\sigma_{los}$(km s$^{-1}$) & 645$\pm$69 & 693$\pm$65\\
$R_{proj}$(Mpc) & 0.75$\pm$0.11 & 0.94$\pm$0.20\\
$M_{vir}$(10$^{14}$ $M_{\odot}$) & 2.2$\pm$0.6 & 3.2$\pm$0.9\\
$r_{200}$(Mpc) & 0.49$\pm$0.05 & 0.52$\pm$0.05\\
$M_{200}$(10$^{14}$ $M_{\odot}$) & 1.4$\pm$0.5 & 1.7$\pm$0.5\\
\\
\hline
\label{table1}
\end{tabular}
\end{scriptsize}
\end{center}
\end{table}

\subsection{Gas Velocity Dispersion} \label{vd}

In order to measure the rest-frame integrated gas velocity dispersion, we use the sigma of the fitted Gaussian function to either the H$\alpha$ or [O{\sc iii}]$\lambda$5007 emission lines. However, we need to subtract in quadrature the instrument velocity dispersion due to instrument resolution. Therefore, the rest-frame integrated velocity dispersion in units of $kms^{-1}$ is calculated as $\sigma$=$\frac{c}{\lambda}\sqrt{\sigma_{m}^{2}-\sigma_{inst}^{2}}$ where $\lambda$ is the redshifted wavelength in \AA ngstrom (i.e.; $\lambda$=$\lambda_{0}(1+z)$ where $\lambda_{0}$ is the rest-frame wavelength of the line of interest), $\sigma_{m}$ is the measured velocity dispersion in \AA ngstrom and $\sigma_{inst}$ is the measured instrument resolution in \AA ngstrom. 

\begin{figure*}
\centering
\includegraphics[width=7.0in]{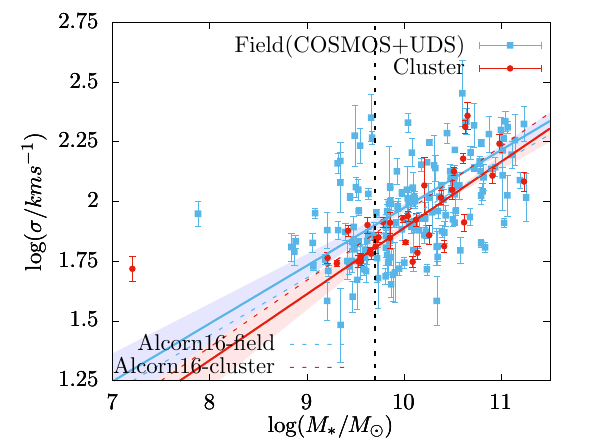}
\caption{Relation between integrated gas velocity dispersion and stellar mass for protocluster (red points) and field (blue points) galaxies. The best-fit lines and their uncertainties are shown with red (protocluster) and blue (field) solid lines and shaded regions, respectively. Vertical black dashed line indicates the stellar mass completeness limit. For comparison, the best-fit lines from \cite{Alcorn16} are plotted as dashed red and blue lines as well. Velocity dispersion increases with stellar mass. However, we find that on average, protocluster galaxies have $\sim$0.08 dex lower velocity dispersions (at $\sim$1.8$\sigma$ level) and a factor of $\sim$2 lower dispersions around the best-fit lines than their field counterparts.  
}
\label{fig:sigma}
\end{figure*}

We use the width of the bright sky lines or wavelength-calibration lamp lines in the vicinity of H$\alpha$ or [O{\sc iii}]$\lambda$5007 lines to estimate the instrument velocity dispersion. We estimate $\sigma_{inst}$=2.5(1.9)\AA \ in the vicinity of H$\alpha$([O{\sc iii}]$\lambda$5007) lines. To check for systematics in measuring velocity dispersions, we use data that have both $H$- and $K$-band measurements. We find a very good agreement between the measured velocity dispersions using H$\alpha$ and [O{\sc iii}]$\lambda$5007 lines. For these sources, the more significant measurement of the two is selected as the final measurement.

We note that the estimated gas velocity dispersion, in addition to dynamics, has some other contribution due to e.g. uncertainty principle, thermal broadening (at a typical temperature of T$\sim$ 10$^{4}$ K for HII regions, it is $\sim$ 10 $kms^{-1}$ for the Hydrogen gas), mergers, turbulent motion inside HII regions (typically $\sim$ 20 $kms^{-1}$; \citealp{Shields90}) etc. We do not correct for these smaller contributors and the reported velocity dispersion is the integrated combination of all of these factors.   

\begin{figure*}
\centering
\includegraphics[width=7.0in]{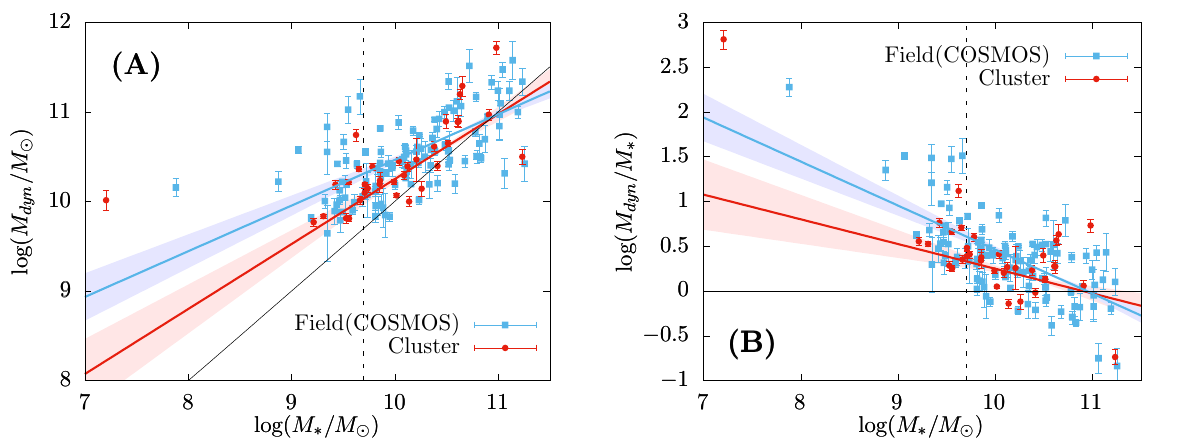}
\caption{
Left panel --- (A) Dynamical mass vs. stellar mass for protocluster (red points) and field (blue points) galaxies. The best-fit lines and their marginal uncertainties are presented with solid red (protocluster) and blue (field) lines. Vertical black dashed line shows the stellar mass completeness limit. Solid black line shows a one-to-one relation. We find that dynamical mass increases with stellar mass. However, on average, protocluster galaxies have $\sim$0.2 dex lower dynamical masses (at $\sim$2.2$\sigma$ level) and $\sim$3 times lower dispersions around the best-fit lines than the field galaxies. Right panel --- (B) Dynamical-to-stellar mass ratio as a function of stellar mass. $M_{dyn}$/$M_{*}$ ratio decreases with increasing stellar mass but protocluster galaxies show $\sim$0.2 dex lower $M_{dyn}$/$M_{*}$ ratio than the field on average.
}
\label{fig:Mdyn}
\end{figure*}

Fig \ref{fig:sigma} shows the velocity dispersion as a function of stellar mass for protocluster and field galaxies. We find that on average, velocity dispersion increases with increasing stellar mass. However, visually we see that both protocluster and field samples follow similar trends. We perform a two dimensional K-S test \citep{Fasano87} to determine if protocluster and field galaxies on the 2D log$\sigma$-log($M_{*}$) plane are drawn from the same parent distribution. We derive a p-value of p=0.09, implying that it is unlikely (at a p=0.09 level) that protocluster and field samples are drawn from different parent 2D distributions. However, we note that the K-S test does not take measurement uncertainties into account. 

For galaxies above the stellar mass limit (log($M_{*}$/$M_{\odot}$)$\geqslant$9.7), we perform a linear fit of the form log$\sigma$=$a$(log($M_{*}$/$M_{\odot}$)-10)+$b$ to protocluster and field samples taking into account the uncertainties in velocity dispersions. To mitigate the role of outliers, we iteratively perform the linear-fit modeling three times, each time removing points whose velocity dispersions are more than three times the median absolute deviation from the best-fit line in each run. We find the following best-fit parameters for protocluster ($a$=0.277$\pm$0.049, $b$=1.889$\pm$0.017) and field ($a$=0.242$\pm$0.034, $b$=1.973$\pm$0.015) samples. The median absolute deviation from the best-fit line after removing the outliers is $\sim$0.054(0.091) dex for protocluster(field) galaxies. This indicates that the dispersion around the average relation is a factor $\sim$ 2 smaller for protocluster galaxies.    

The slope difference between the protocluster and field best-fit relations is consistent within uncertainties. To quantify the slope difference, we performed a t-test, finding one-sided(two-sided) p-value of p=0.494(0.988). This implies that the slope difference is not significant. We find that on average, protocluster galaxies have $\sim$0.08 dex lower velocity dispersions per $M_{*}$ than their field counterparts. However, this difference is of the order of the typical dispersion of the points around the best-fit lines. To quantify this, we perform a t-test. The intercept difference is significant at a $\sim$1.8$\sigma$ level (one-sided p=0.034).

\cite{Alcorn16} performed a linear-fit to the log$\sigma$-log($M_{*}$) relation for a sample of cluster and field galaxies at $z\sim$2 finding consistent results between cluster and field best-fit parameters. In Fig. \ref{fig:sigma}, we compare their best-fit line with ours, finding good agreement between the two works within uncertainties. However, we find tentative evidence ($\sim$1.8$\sigma$) for protocluster galaxies to have lower velocity dispersions than the field systems at fixed stellar masses. 
 
\subsection{Dynamical Mass and Dynamical-to-Stellar Mass Ratio} \label{dm}

We estimate the dynamical mass within the circularized effective radius as $M_{dyn}$=$\frac{\beta \sigma^{2} R_{cir}}{G}$ where $\beta$ is a parameter that depends on the mass distribution within a galaxy, $\sigma$ is the velocity dispersion, $R_{cir}$ is the circularized effective radius and G is the gravitational constant. $\beta$ value ranges between $\sim$ 2-10 in the literature (e.g., \citealp{Erb06b,Beifiori14,Alcorn16}). Here, we use $\beta$=10.
We estimate $M_{dyn}$ uncertainties via 10000 bootstrap resamples using uncertainties in velocity dispersions only.

Figure \ref{fig:Mdyn} (A) shows the relation between $M_{dyn}$ and $M_{*}$ for protocluster and field samples. It is clear that $M_{dyn}$ increases with increasing stellar mass. Visually, this relation seems to be similar for both protocluster and field galaxies. To quantify this, we perform a 2D K-S test. The difference is at p=0.07 level, meaning that it is only $<$7\% probable that the observed difference in log($M_{dyn}$)-log($M_{*}$) distributions in different environments is due to chance.

Similar to Section \ref{vd}, for log($M_{*}$/$M_{\odot}$)$\geqslant$9.7 galaxies, we perform a linear regression fit of the form log($M_{dyn}$/$M_{\odot}$)=$a$(log($M_{*}$/$M_{\odot}$)-10)+$b$ considering uncertainties in $M_{dyn}$ and removing outliers using a sigma-clipping method. The result of the fit is ($a$=0.725$\pm$0.118, $b$=10.249$\pm$0.037) and ($a$=0.509$\pm$0.076, $b$=10.463$\pm$0.036) for protocluster and field samples, respectively. The median absolute deviation from the best-fit line after removing the outliers is $\sim$0.066(0.178) dex for protocluster(field) galaxies, $\sim$3 times smaller for protocluster systems. 

We find no statistically significant difference between the slopes. The one-tailed(two-tailed) p-value for a t-test is p=0.473(0.945). The intercept difference is $\sim$0.2 dex. This indicates that on average, protocluster galaxies have $\sim$ 0.2 dex lower $M_{dyn}$ than the field, particularly for less massive systems as seen in Figure \ref{fig:Mdyn}. A t-test analysis shows that the difference is statistically significant at a $\sim$2.2$\sigma$ level (one-sided p-value=0.014). To make sure that this difference is not caused by different stellar mass distributions between protocluster and field galaxies, we performed a 1D KS test on stellar masses. We find a KS p-value=0.58, indicating it is very unlikely the stellar masses in these two samples are drawn from different parent distributions. We also note that the dispersion around this best-fit relation is relatively large ($\sim$0.2 dex), especially for field galaxies which is similar to the $M_{dyn}$ difference ($\sim$0.2 dex) in these two environments. 

Figure \ref{fig:Mdyn} (B) better shows the $M_{dyn}$ difference between protocluster and field galaxies at fixed $M_{*}$. We find an anti-correlation between stellar-to-dynamical mass ratio ($M_{dyn}$/$M_{*}$) and stellar mass, with less massive galaxies having larger $M_{dyn}$/$M_{*}$ per stellar mass. More importantly, we find that on average, protocluster galaxies have $\sim$ 0.2 dex lower dynamical-to-stellar mass ratio at fixed stellar masses than their field counterpart, particularly at lower $M_{*}$ values. The result of the best-fit line to log($M_{dyn}$/$M_{*}$) vs. log($M_{*}$) relation is ($a$=-0.275$\pm$0.118, $b$=0.249$\pm$0.037) for the protocluster and ($a$=-0.490$\pm$0.076, $b$=0.463$\pm$0.036) for the field galaxies. We also find that the dispersion around this relation is $\sim$3 times larger for field galaxies (MAD value after removing the outliers is $\sim$0.066 for protocluster vs. $\sim$0.178 for the field). 

\section{Discussion} \label{dis}

\cite{Cucciati18} reported the discovery of a massive $\sim$ 60$\times$60$\times$150 Mpc$^{3}$ (comoving) LSS at $z\sim$2.45 in the COSMOS field, dubbed Hyperion. Hyperion has seven spectroscopic confirmed overdensities with masses ranging $\sim$ (0.1-2.7)$\times$10$^{14}$ $M_{\odot}$. We believe the LSS presented in Figure \ref{fig:map} (A) (also \citealp{Darvish20}) is similar to Hyperion but at marginally lower redshifts. So far, it has at least three spectroscopically-confirmed density peaks and it is extended over a comoving volume of $\sim$ 40$\times$40$\times$180 Mpc$^{3}$. \cite{Ata20} performed constrained simulations using spectroscopic data in the COSMOS field at $z\sim$2.3. They found massive halos at the positions of the Hyperion and CC2.2 in their simulated realizations. In 84\% of their realizations, they found a cluster with a present-time mass of $M(z=0)$=(4.2$\pm$1.9)$\times$10$^{14}$ $h^{-1}M_{\odot}$ that would finally form out of an overdensity close to CC2.2. They found the mean halo mass was less massive than what was estimated for CC2.2 ($M(z=0)$=9.2$\times$10$^{14}$ $M_{\odot}$) in \cite{Darvish20} but the difference was only at $\sim$1.44$\sigma$ level which could be due to differences in methods and selections used.

If enough time has passed since the formation of the protoclusters, member galaxies effectively interact gravitationally with one another and the potential well of the dense environment of the protocluster. Hence, they eventually lose their mass as a result of these frequent gravitational interactions and galaxy harassment. However, one expects the dark matter component of a galaxy to undergo a larger mass loss than the stellar component. This is because the stellar component is more gravitationally bound to the parent galaxy as the majority of the stellar mass is located in the central part of the galaxy halo. Therefore, one expects a smaller dynamical-to-stellar mass ratio for galaxies in dynamically evolved dense environments than the field. This is in full agreement with what we find in Section \ref{dm}, with protocluster galaxies having $\sim$0.2 dex lower dynamical masses than the field sample at fixed $M_{*}$. 

The environmental dependence of the dynamical measures for star-forming galaxies can also be used to place constraints on the formation time of our protoclusters. In order for protocluster galaxies to gravitationally feel the potential well of their dense environment as well as other member galaxies, at least one dynamical timescale must have passed since the formation of their host environment. We estimate the dynamical timescale using $\tau_{dyn}\sim r_{3d}/\sigma_{3d}$ where $r_{3d}$ is the typical radius of the protocluster and $\sigma_{3d}$ is its total velocity dispersion. Assuming $r_{3d}\sim R_{proj}$ and a spherical symmetry for our protoclusters (i.e.; $\sigma_{3d}$=$\sqrt{3}\sigma_{los}$) we obtain $\tau_{dyn}\sim$0.6-0.7 Gyr. This means that our protoclusters must have formed at least at $z\gtrsim$2.8-2.9. In practice, a few dynamical timescales must have passed, placing the formation time of our protoclusters at even higher redshifts.        

\cite{Alcorn16} investigated the role of environment on the kinemtics of a sample of $\sim$75 star-forming galaxies from the ZFOURGE survey at $z\sim$2. They found no statistically significant differences between cluster and field log($M_{*}$/$M_{\odot}$)=9-11 star-forming galaxies. Here, we found subtle and tentative differences between the kinematics of protocluster and field galaxies at $z\sim$2. 

In Sections \ref{vd} and \ref{dm}, we found that protocluster SF galaxies have a factor of $\sim$2-3 lower scatter around their mean $\sigma$-$M_{*}$, $M_{dyn}$-$M_{*}$, and $M_{dyn}$/$M_{*}$ vs. $M_{*}$ relations compared to similar relations for the field control sample. This could be due to a more uniform and coherent formation for protocluster galaxies than their field counterparts.

In Section \ref{dm}, we used a value of $\beta$=10 for estimating the dynamical masses. Even with this extreme selection, $\sim$ 11(18)\% of cluster(field) galaxies yield seemingly unphysical values for $M_{dyn}$/$M_{*}$ and this situation is worse for more massive galaxies (Fig. \ref{fig:Mdyn} B). This can constrain how mass is distributed (which is related to the $\beta$ value) within $z\sim$2 star-forming galaxies as a larger $\beta$ value is preferred. However, the unphysical values could also be due to uncertainties in the estimated stellar and dynamical masses. The uncertainties presented in fig. \ref{fig:Mdyn} (B) do not contain those of the $M_{*}$ and size measurements and the real uncertainties are larger. Moreover, the nebular gas dynamics may not be a full representative of the dynamical mass measurement as the star-forming regions could be mostly localized to inner parts of a galaxy. We also note that $M_{*}$ and $M_{dyn}$ are measured in different apertures. According to Fig. \ref{fig:Mdyn} (B), the situation is worse for more massive galaxies, indicating that for less massive systems, much of their dynamical mass is constrained within the central regions. 
 
We note that galaxy interaction can trigger starbursts or AGN activity (e.g., \citealp{Bergvall03}) and the removal of AGN and mergers in their final stages can somehow bias our dynamical analysis of galaxies. Therefore, our results apply only to normal star-forming galaxies experiencing galaxy harassment and interactions involving high-speed flybys.      

\section{summary}

In this paper, we spectroscopically confirm a new protocluster at $z$=2.24430 in the COSMOS field using Keck/MOSFIRE, dubbed CC2.2B. The centroid of CC2.2B is only separated by $\sim$5.5 Mpc(angular) and $\sim$16 comoving Mpc(radial) from that of CC2.2A, originally presented in \cite{Darvish20}. CC2.2B has a velocity dispersion and estimated total mass of $\sigma_{los}$=693$\pm$65 km s$^{-1}$ and $M$=($\sim$2-3)$\times$10$^{14}$ $M_{\odot}$, respectively, similar to those of CC2.2A (Table \ref{table1}). These two protoclusters are likely still in the merging process. We combine CC2.2A and CC2.2B data to investigate the role of high-$z$ protocluster environments on the kinematics of galaxies compared to a similarly selected sample in the field. We use the integrated gas velocity dispersion, estimated by measuring the width of nebular H$\alpha$ or [O{\sc iii}]$\lambda$5007 emission lines, as a measure of kinematics of galaxies. Combined with HST size measurement, we compute the dynamical mass of our sample galaxies within their circularized effective radius. We find that:

\begin{enumerate}

\item Gas velocity dispersion increases with stellar mass. However, we find that on average, protocluster galaxies have $\sim$0.1 dex ($\sim$1.8$\sigma$) smaller velocity dispersions and $\sim$2 times lower scatter around the mean relation than the field galaxies.

\item For both protocluster and field galaxies, the dynamical mass increases with increasing stellar mass. However, on average, protocluster galaxies have $\sim$0.2 dex ($\sim$2.2$\sigma$) lower velocity dispersions and a factor of $\sim$3 lower dispersions around the mean relation than their field counterparts.

\item We see an anti-correlation between dynamical-to-stellar mass ratio vs. stellar mass, with less massive galaxies having a higher $M_{dyn}$/$M_{*}$ ratio. We find a $\sim$0.2 dex lower $M_{dyn}$/$M_{*}$ ratio and a smaller scatter around the mean relation ($\sim$3$\times$) for protocluster galaxies than the field sample.  

\end{enumerate}

We suggest galaxy-galaxy interactions and galaxy harassment as a physical explanation for these subtle differences between protocluster and field galaxies at $z\sim$2.2. We also suggest that protocluster SF galaxies likely have a more uniform formation time than the field sample. Our protoclusters are in the footprint of the JWST COSMOS-Web survey \citep{Casey22}. The survey will soon provide deep and high resolution (5$\sigma$ point source depths of $\sim$27.5-28.2 magnitudes) NIRCam imaging data in four bands. This will allow us to make accurate size measurements for our sample in near future.   
 
\section*{acknowledgements}

The observations presented herein were obtained at the W. M. Keck Observatory, which is operated as a scientific partnership among the California Institute of Technology, the University of California, and the National Aeronautics and Space Administration. The Observatory was made possible by the generous financial support of the W. M. Keck Foundation. The authors would like to recognize and acknowledge the very prominent cultural role and reverence that the summit of Mauna Kea has always had within the indigenous Hawaiian community. We are fortunate to have the opportunity to perform observations from this mountain.

\bibliographystyle{aasjournal} 
\bibliography{references}

\end{document}